\documentclass[11pt,twoside]{article}

\usepackage{asp2004}
\usepackage{epsf}
\usepackage{psfig}
\usepackage{lscape}

\begin{document}
\title{Stellar Populations in the Magellanic Clouds}    
\author{Nino Panagia$^1$, Guido De Marchi$^2$, Martino Romaniello$^3$}   
\affil{(1) Space Telescope Science Institute, 3700 San Martin Drive, Baltimore, MD 21218, USA; also,
INAF-HQ, Via del Parco Mellini 84, 00136 Rome, Italy; and Supernova
Ltd., OYV \#131, Northsound
Road, Virgin Gorda, British Virgin Islands \\(2)
ESA/ESTEC, Keplerlaan 1, 2200 AG Noordwijk, The Netherlands \\(3) ESO,
Karl-Scharzschild-Strasse 2, 85748 Garching-bei-M\"unchen,
Germany}    

\begin{abstract} 
We present the first results of our study of stellar populations in the
Large and Small Magellanic Clouds based on multiband WFPC2 observations
of ``random" fields taken as part of the ``pure parallel" program
carried out by the Space Telescope Science Institute as a service to the
community.
\end{abstract}


We have started a study of the stellar populations in the
Large and Small Magellanic Clouds using data (U, B, V, I and H$\alpha$ 
imaging) collected as part of the WFPC2 Pure Parallel Program
(Wadadekar et al. 2006). 
We have  considered a number of fields in both Magellanic Clouds which
comprise stellar populations with  ages ranging from a few Myr to a few
Gyr and in which pre-main-sequence objects are clearly identified. In
addition to the properties of these populations, we have studied the
characteristics of the dust present in these regions. 

Using the properties of the red giants of the so-called ``red clump"
(RC) in the various
colour-magnitude diagrams, we have determined the extinction law in each
field (Panagia \& De Marchi 2005). 
We find that not only the amount of extinction but also its wavelength
dependence varies considerably from site to site, indicating that the
physical properties of the absorbing dust are not uniform across the
Clouds (see Figure 1). 
This analysis also allows us to determine the relative three-dimensional
distribution of the  different populations as well as their location
with respect to the  absobing dust clouds. 

Correcting for the appropriate extinction for each field {\it
individually}, and following  the procedures developed by Panagia et al.
(2000), Romaniello et al. (2002), and Romaniello, Robberto \& Panagia
(2004), we determine the physical parameters, namely effective
temperature and luminosity, of all stars present in the field (De Marchi
et al. 2005).  Table 1 illustrates some of the properties of the stellar
populations for the fields we have analyzed so far.

We find that,  among young stars, those less  massive than 2 solar
masses are spatially more spread over each field  than massive B type
stars, whereas old stars are quite uniformly distributed. The data also
reveal that young stars, which appear to be spatially associated with
the nebular emission, are physically confined within a small range of
distances, whereas old stars are uniformly distributed at all depths
along the line of sight. 
An important corollary of our investigation is that unaccounted patchy
absorption or a variable extinction law are likely to contribute
significantly to the present discrepancies between various distance
indicators for the Magellanic Clouds.



\begin{figure}[!ht]
\vspace{2.9cm}
\plotfiddle{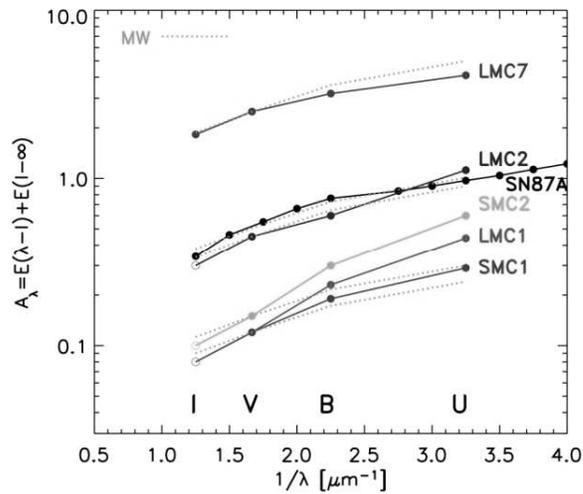}{3.8cm}{0}{35}{35}{-120}{-35}
\caption{Reddening curves for several fields in the LMC and SMC compared
with the standard Milky Way extinction law (dotted lines).}
\end{figure}

\vskip .1cm

\begin{table}[!ht]
\caption{Properties of Stellar Populations in several LMC and SMC Fields}
\begin{center}
{\small
\begin{tabular}{lccccccc}
\tableline
\noalign{\smallskip}
~Field & E(V-I) & $<$A$_V>$ & V$_{min}$ & M$_{max}$ & T$_{min}$ & T$_{RC}$
&Z$_{RC}$\\
  & range & & brightest  & highest    & youngest  & min age & metallicity \\
  &       & & young      & mass       & blue      & RC      & range RC\\
  &       & & star       & star       & star      &         &          \\
  &       & &            & [M$_\odot$]& [Myr]     & [Gyr]   & [Z$_\odot$] \\
\noalign{\smallskip}
\tableline
\noalign{\smallskip}
LMC1  &  0.00-0.06 & 0.14 & 15.5 &  5 & 100 & 1.4 & 0.2-0.4  \\
LMC2  &  0.07-0.20 & 0.50 & 16.1 &  6 &  70 & 1.0 & 0.2-0.6  \\
LMC7  &  0.30-0.70 & 1.95 & 13.8 & 36 &   4 & 0.5 & 0.2-0.6  \\
SN87A &  0.20-0.50 & 1.23 & 13.3 & 14 &  16 & 0.5 & 0.2-0.4  \\
SMC1  &  0.04-0.08 & 0.21 & 15.8 &  9 &  30 & 1.4 & $<$0.2   \\
SMC2  &  0.08-0.16 & 0.42 & 16.5 &  7 &  50 & 1.4 & $<$0.2   \\
\noalign{\smallskip}
\tableline
\end{tabular}
}
\end{center}
\end{table}

\end{document}